\documentclass[letterpaper]{article}

\usepackage{graphicx}
\graphicspath{{./Fig/}}
\usepackage{color}
\usepackage{placeins}
\usepackage{float}
\usepackage{tabularx,colortbl}
\usepackage{amsmath}
\usepackage{amsfonts}
\usepackage{caption}
\usepackage{soul}
\usepackage{authblk}
\usepackage{amssymb}
\usepackage{amsmath}
\usepackage{amsthm}
\usepackage{amsfonts}
\usepackage[sort&compress,round,colon,authoryear]{natbib}
\usepackage{authblk}
\usepackage{indentfirst}
\usepackage{subfig}
\usepackage{mathtools}
\usepackage{bm}
\usepackage{mathrsfs}
\usepackage{fancyhdr}
\usepackage{cite}
\usepackage{enumerate}
\usepackage{booktabs}
\usepackage{colortbl}
\usepackage{multicol}
\usepackage{multirow}
\usepackage{cancel,soul}
\setstcolor{red}
\usepackage{xcolor}
\usepackage{nth}
\usepackage{url}

\usepackage{stfloats}
\usepackage[font=small,skip=0pt]{caption}
\usepackage[boxed,linesnumbered]{algorithm2e}
\usepackage{listings}
\providecommand{\keywords}[1]{\textbf{\textit{Keywords---}} #1}

\newcommand{\Ex}{\mathbb{E}}

\newcommand{\cF}{\mathcal{F}}

\title{Microseismic events enhancement and detection in sensor arrays using autocorrelation based filtering}
\author[1]{Entao Liu\thanks{liuentao@gmail.com}}
\author[1]{Lijun Zhu}
\author[1]{Anupama Govinda Raj}
\author[1]{James H. McClellan}
\author[2]{Abdullatif Al-Shuhail}
\author[2]{SanLinn I. Kaka}
\author[2]{Naveed Iqbal}
\affil[1]{CeGP at Georgia Institute of Technology}
\affil[2]{CeGP at KFUPM}

\begin{document}
\maketitle

\renewcommand{\thefootnote}{\fnsymbol{footnote}}

\begin{abstract}
Passive microseismic data are commonly buried in noise, which presents a significant challenge for signal detection and recovery. 
For recordings from a surface sensor array where each trace contains a time-delayed arrival from the event, we propose an autocorrelation-based stacking method that designs a denoising filter from all the traces, as well as a multi-channel detection scheme. This approach circumvents the issue of time aligning the traces prior to stacking because every trace's autocorrelation is centered at zero in the lag domain. 
The effect of white noise is concentrated near zero lag, so the filter design requires a predictable adjustment of the zero-lag value. Truncation of the autocorrelation is employed to smooth the impulse response of the denoising filter. In order to extend the applicability of the algorithm, we also propose a noise prewhitening scheme that addresses cases with colored noise.
The simplicity and robustness of this method are validated with synthetic and real seismic traces.   
\end{abstract}


\newpage

\section{Introduction}
Microseismic monitoring is of great interest for its capability of providing valuable information for many oil and gas related applications, such as hydraulic fracturing monitoring, unconventional reservoir characterization, and $\text{CO}_2$ sequestration (\citealt{DunEis2010}; \citealt*{MaxWhiAO2004}; \citealt{PhiRutAO2002};  \citealt{War2009}).
In practice, microseismic monitoring is performed using downhole, surface, or near surface arrays (\citealt{DunEis2010}). Recently, there is a preference for surface arrays which can be deployed more economically and efficiently. However, recorded surface microseismic signals are much noisier than downhole data, because surface or near surface arrays are susceptible to both coherent and incoherent noise.

Consequently, it is challenging to extract useful information from the recorded microseismic data. Typically, the signal-to-noise ratio (SNR) of recorded microseismic data is rather low, especially for the data collected using surface arrays. The noise adversely affects the accuracy of many microseismic analyses, such as P-wave and S-wave arrivals picking, event detection and localization, and focal mechanism inversion (\citealt{SabVel2010}; \citealt*{ZhaTianAO2014}; \citealt*{ZhuLiuAO2015}). In fact, the majority of microseismic events induced by hydraulic fracturing has a typical moment magnitude $M_W<-1$ (\citealt{SonWarAO2014}). 

We consider the case of a sensor array that records multiple traces, and assume that the microseismic signal of interest is present in all the traces.
Therefore, stacking is one of the primary techniques to consider for improving SNR, as it is well known for seismic applications (\citealt{Yil2001}). However, for microseismic data time alignment of the traces, which is the prerequisite of stacking, is generally unknown. Thus, researchers have developed algorithms based on cross-correlation to find the relative time delays between traces (\citealt*{AlkaAO2013}; \citealt{GreZha2012}). In contrast to the case of an active seismic where the source generates a controllable active wavelet, the wavelet of a microseismic event is unknown, although some empirical knowledge of the frequency domain characteristics may be available. The cross-correlations are usually computed from noisy traces rather than from a clean signal template. Therefore, the maximal value of the cross-correlation may occur at an incorrect lag because of the noise. 
To bypass this bottleneck, we propose denoising and detection schemes using stacked autocorrelograms, which are automatically 
aligned at the zero lag. The simplicity and robustness of the proposed schemes are demonstrated by using both synthetic and real data. In the literature, the stacked cross-correlation and autocorrelation have already shown promising results in different scenarios (\citealt{LiuZhu2016}; \citealt{WapDraAO2010,WapSlo2010}; \citealt{ZhaTianAO2014}). 

\section{Denoising}
\subsection*{Problem formulation}
Assume the sensor array for microseismic monitoring has $N$ channels. Each recorded trace contains a microseismic waveform contaminated by noise, i.e.
\begin{equation}\label{model}
x_i(t) = a_i s_i(t)+n_i(t),\ \ \ i=1,\ldots,N,
\end{equation}
where $a_i$ are amplitude scaling factors, and $n_i(t)$ is zero mean additive white Gaussian noise (AWGN) with variance $\sigma^2$.  For simplicity, we first consider only white noise whose power spectrum is flat; the colored noise scenario, whose power density has different amplitudes for different frequencies, will be discussed in a later section.

Theoretically, seismic traces received on different sensors are convolutions of the same source wavelet with Green's functions which are determined by the real Earth and source and sensor locations (\citealt{AkiRic2009,SteWys2002}). Nonetheless, a high resemblance between different traces originated by the same event and collected by spatially close sensors is usually observed (\citealt{ArrEis2006}). Therefore, it is reasonable to assume that $s_i(t)$ are all the same waveform but with different time delays, i.e., $s_i(t) = s(t-\tau_i)$.
In addition, we assume the $n_i(t)$ are uncorrelated with the waveform $s(t)$ and independent of each other; uncorrelated noise on different channels is a common assumption for the validity of any stacking technique. 

The processing is performed on digital signals,  sampled  at a rate $f_s = 1/\Delta t$, so that  the time variable $t$ becomes $t_l=t_0+(l-1)\Delta t$ for $l=1,\ldots,L$.
On each trace, we consider a finite time window of data which contains $L$ time samples, so the signals can be written as 
\begin{equation}
x_i[l]=a_i s_i[l] +  n_i[l], \qquad l=1,\ldots,L
\end{equation} 
where $x_i[l]=x_i(t_l)$. 
In this work, we define the SNR of the $i$-th channel as the ratio of the signal energy to the AWGN energy, i.e., 
\begin{equation}\label{SNR}
\text{SNR}_i=10\log_{10}\left(\frac{a_i^2\|s_i\|_2^2}{\|n_i\|_2^2}\right).
\end{equation}
As reference, we use the peak-signal-to-noise ratio (PSNR) as well
\begin{equation}\label{PSNR}
\text{PSNR}_i=10\log_{10}\left(\frac{a^2_i P^2_i}{\sigma^2}\right),
\end{equation}
where $P_i$ is the peak value of $s_i(t)$. When the wavelet duration is short, the PSNR is more intuitive than SNR, because its value is not affected by the signal length.  

\subsection{Denoising Filter Design}
%

The proposed approach is implemented with the following three steps:
\begin{enumerate}
	\item[1. ] Compute the autocorrelation function (ACF) of each trace (denoted by $\star$) and then stack the ACFs,
	\begin{equation}
	r_s[\tau] = \frac{1}{N}\sum_{i=1}^N (x_i \star x_i)[\tau], 
	\end{equation}
	where $\tau=-L+1,\ldots,0,\ldots,L-1$ is the lag index of the ACF.
	\item[2. ]
	Define the denoising filter's impulse response as a windowed version of the modified ACF, $f[\tau]=\hat{r}_s[\tau]w_d[\tau]$, where
	\begin{equation}\label{filterDesign}
	\hat{r}_s[\tau] = \left\{ \begin{array}{cl}
	\frac{1}{2}\big(r_s[-1]+r_s[1]\big)&\mbox{ if $\tau = 0$} \\
	r_s[\tau] &\mbox{ otherwise}
	\end{array} \right.
	\end{equation}
	The zero-lag value of the ACF is replaced with the average of its neighboring values, $r_s[1]$ and $r_s[-1]$.
	The justification for this change is that additive white noise has an ACF that is $L\sigma^2 \delta[\tau]$, where $\delta[\tau]$ is the discrete Dirac delta impulse function and only gives nonzero value at $\tau=0$.
	Thus the ACF $r_s[\tau]$ has a large peak at $\tau=0$ which needs to be reduced by $L\sigma^2$; the average of the neighboring values provides an estimate of the correct zero-lag value of the signal-only  ACF.
	\item[3. ] 
	A truncation window $w_d[\tau]$ is then applied to the zero lag region, if necessary, so that the negligible values in the filter (away from $\tau=0$) will be eliminated. A proper truncation window that shortens the filter length will improve the computational efficiency as well. Various truncation windows are available, however we adopt a simple triangle window
	\begin{equation}\label{truncwindow}
	w_d[\tau] = \begin{cases}
	1 - |\tau|/d &\text{if } |\tau| \leq d \\
	0 &\mbox{ otherwise}
	\end{cases}
	\end{equation}
	where $2d+1$ is the length of the truncation window. 
	\item[4. ] Convolve $f[\tau]$ with each noisy trace in the collection. The result of these convolutions provides the $N$ denoised traces $\hat{x}_i[l]=(f \ast x_i) [l]$ for $i=1,2,\ldots,N$.
\end{enumerate}

\begin{figure}[htbp]
	\begin{center}
		\includegraphics[width=0.7\textwidth]{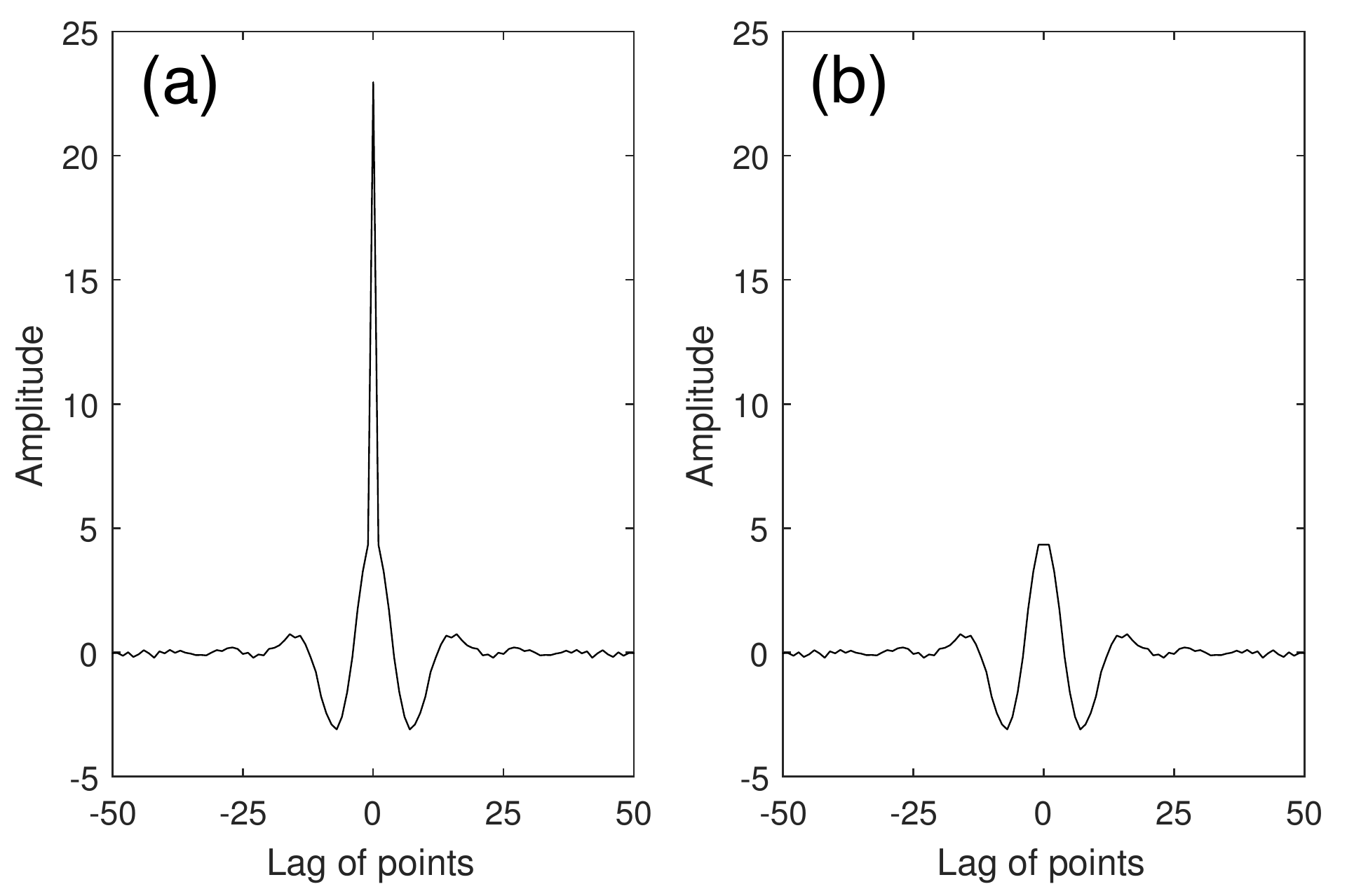}
		\caption{(a) Stacked autocorrelation (b) denoising filter. $L=200$ and $\sigma=0.3$, so the peak height removed is approximately $L\sigma^2=18$.}
		\label{fig:autoVSfilter}
	\end{center}
\end{figure}

In Figure \ref{fig:autoVSfilter}, we show the stacked autocorrelation $r_s[\tau]$ and the filter $f[\tau]$ based on it for the case of a 30-Hz Ricker wavelet sampled at $f_s = 500$\,Hz, which is also treated in the denoising example for synthetic data in Section \ref{sec:synthDataRicker}.  The spiky peak in $r_s[\tau]$ at zero lag is removed by the average operation  \eqref{filterDesign} to obtain $f[\tau]$. 

Compared with a cross-correlation based method, this new autocorrelation-based approach has approximately the same computational burden, since computing one ACF or one cross-correlation function (CCF) is $\mathcal{O}(L\log_2(L))$. In fact, some overhead is removed by avoiding the search for the maximal value needed for alignment.

\subsection{Frequency Response Analysis}
In order to evaluate the performance of the designed filter, we
examine the frequency response of the filter $f[\tau]$ since the analysis of the filter performance is easier to conduct in the frequency domain. The Fourier transform $(\mathcal{F})$ of the ACF $r_s(\tau)$ is
\begin{align}\label{FFT}
R_s(e^{j\omega})&=\mathcal{F}  \{ r_s(\tau)\}\nonumber\\
& =\frac{1}{N}\sum_{i=1}^N X_i(e^{j\omega})\overline{X_i(e^{j\omega})}\nonumber\\
& =  \frac{1}{N}\sum_{i=1}^N  (a_i^2 |S_i(e^{j\omega})|^2 + |N_i(e^{j\omega})|^2 + a_i S_i(e^{j\omega}) \overline{N_i(e^{j\omega})}+a_i N_i (e^{j\omega})\overline{S_i(e^{j\omega})})\nonumber\\
& \approx |S(e^{j\omega})|^2 \frac{1}{N}\sum_{i=1}^N  a_i^2 + \frac{1}{N}\sum_{i=1}^N  |N_i(e^{j\omega})|^2,
\end{align}
where the last line is derived from the assumption that the signal $s(t)$ and noise $n_i(t)$ are uncorrelated. The two terms in  \eqref{FFT} are scaled versions of the power spectrum of the signal (i.e., magnitude squared),  and the power spectrum of the noise which is a random variable at each frequency $\omega$.
If we assume $a_i=1$ for all $i$, then there is no scaling of the first term. The second term is the expected value of the noise power spectrum which for white noise is equal to $L\sigma^2$ for all frequencies.


We can now justify the definition of $\hat{r}_s[\tau]$ in equation \eqref{filterDesign}.  Compared with the $r_s(\tau)$, the designed filter is modified slightly from the autocorrelation by removing the peak at zero lag. However, the frequency response is beneficially manipulated for the denoising purpose. For ease of exposition, we assume $a_i=1$ for all $i$. When the sampling frequency is high enough and noise is white $f[0]=\frac{1}{2}(r_s[-1]+r_s[1])=r_s[-1]$ turns out to be an accurate approximation of the energy of $s(t)$. We then can decompose $r_s[0]$ into two parts
\begin{equation}
r_s[0]=\hat{r}_s[0]+p.
\end{equation}
Again, because signal $s(t)$ and noise $n_i(t)$ are uncorrelated, 
\begin{equation}\label{r0}
r_s[0]=\frac{1}{N}\sum_{i=1}^N ( \|x_i\|^2_2) = \|s\|_2^2 + \frac{1}{N}\sum_{i=1}^N\| n_i\|_2^2,
\end{equation}
and
\begin{equation}\label{f0}
\hat{r}_s[0]\approx \|s\|_2^2.
\end{equation}
Subtracting equation \eqref{f0} from \eqref{r0}, we obtain 
\begin{equation}\label{parseval}
p\approx \frac{1}{N}\sum_{i=1}^N\| n_i\|_2^2=\frac{1}{NL}\sum_{i=1}^N\| N_i(e^{j\omega})\|_2^2,
\end{equation}
by Parseval's theorem. Therefore, $p$ is approximately the average energy of the noise over all traces. 

We reformulate $r_s[\tau]$ as
\begin{equation}
r_s[\tau]=\hat{r}_s[\tau]+p\delta[\tau].
\end{equation} 
For a certain frequency $\omega$, the frequency response of the filter $\hat{r}_s[\tau]$ is expected to be the power of the signal at $\omega$ when the noise is white,
\begin{equation}\label{FRfilter}
\begin{aligned}
\Ex[\mathcal{F}\{\hat{r}_s[\tau]\}] &= \Ex [\mathcal{F}\{r_s[\tau]-p\delta[\tau]\}]\\
&=\Ex[\mathcal{F}\{r_s[\tau]\}]-p\\
&\approx \Ex[\frac{1}{N} \sum_{i=1}^N  | X_i(e^{j\omega})|^2]-\frac{1}{NL}\sum_{i=1}^N\| N_i(e^{j\omega})\|_2^2\\
&\approx \Ex [\frac{1}{N}\sum_{i=1}^N  | S_i(e^{j\omega})|^2]+\Ex[\frac{1}{N}\sum_{i=1}^N | N_i(e^{j\omega})|^2] -\frac{1}{NL}\sum_{i=1}^N\| N_i(e^{j\omega})\|_2^2\\
&\approx \Ex [\frac{1}{N}\sum_{i=1}^N  | S_i(e^{j\omega})|^2].
\end{aligned} 
\end{equation}
Here we can see that the proposed filter $\hat{r}_s[\tau]$ (without truncation) shifts the frequency response amplitude of $r_s[\tau]$ down by $p$. The frequency response of the filter is literally an estimation of the power spectrum of the signal. 

Due to the focal mechanism of the seismic source, the waveforms received on the sensor array can have different amplitudes and phases (i.e., the signs of $a_i$). Corrections to the signs are helpful when we stack the traces or perform the cross correlations. Otherwise, stacking waveforms of different signs would cancel each other rather than enhance them. One advantage of ACF-designed filter $f[\tau]$ over cross-correlation based method is that $f[\tau]$ is not affected by the signs of $a_i$. Essentially, stacking autocorrelation is equivalent to stacking squared frequency response amplitude in frequency domain where all $a_i$ are squared as in equation \eqref{FFT}. The dispersion of wave is another phenomenon which is harmful to the stacking techniques in the time domain. However, the stacked autocorrelation is not affected by dispersion, for the same reason. Although the idea of enhancing the signal in the frequency domain is widely used, one obvious advantage of the proposed scheme is that it produces a band pass filter (BPF) that adapts to the received signals instead of using a filter whose pass band is specified {\it a priori}.


Recall the truncation window in equation \eqref{truncwindow} applied to the filter, which expedites the convolution step. In addition, the truncation of the filter introduces a smoothing effect in its frequency response. Usually, the smoothing effect is beneficial to the denoising filter performance.


\subsection{Noise whitening}
One of the basic assumptions behind the proposed denoising scheme is that of additive white noise. In practice, seismic noise is often colored when it is related to the environment, weather, or human activities. 
And hence the autocorrelation of the noise could be far from an impulse function. So, we would not be able to eliminate the noise's frequency response by manipulating the stacked autocorrelation value only at the zero lag position. To generalize for the colored noise case, we advocate a noise whitening scheme based on linear prediction theory. 

Before discussing the whitening algorithm to flatten the noise spectrum, we state two assumptions: (1) the samples of the colored noise are from a stationary random process, and (2) we have a snapshot of the noise-only data of sufficient time duration. This second assumption would be practical for microseismic monitoring, since the noise-only data can be recorded before the fluid injection activities are started.

Let $v(t)$ be the stationary, colored noise signal. Any wide sense stationary random process can be modeled using an Auto Regressive (AR) model if the order of the model is chosen large enough (\citealt{Kay1988,Mak1975}). Then we can apply a linear prediction filter with coefficients $\{c_k\}$ to predict the next sample using $P$ previous samples
\begin{equation}
\hat{v}[l]=\sum_{k=1}^P c_k v[l-k],
\end{equation}
where $P$ is the order of the prediction filter. 
In general, the prediction is not perfect, so there will be a prediction error sequence 
\begin{equation}
e[l]=v[l]-\hat{v}[l].
\end{equation}
The optimal linear predictive coding (LPC) coefficients minimizes the prediction error power of the $P$-th order linear predictor (\citealt{Kay1988}). This operation has been shown equivalent to maximizing spectral flatness at the output of prediction error filter (\citealt{GarMar1974, MarGar1976}). For an AR($P$) process, the optimal prediction coefficients are the AR($P$) parameters itself. The prediction error filter thus acts as a whitening filter that decorrelates the input AR process to produce white noise as its output. Determining $P$ is the model order selection problem which can be handled by using the fact for a linear predictor of order $q$, satisfying $q>P$ where $P$ is the true order of the AR process, the error power is constant (\citealt{Kay1988}). Even if the order of the AR($P$) process and the order $q$ selected for the linear predictor are not the same, the output prediction error will be white if $q\geq P$.

\begin{figure}[htbp]
	\centering
	\includegraphics[width=0.8\textwidth]{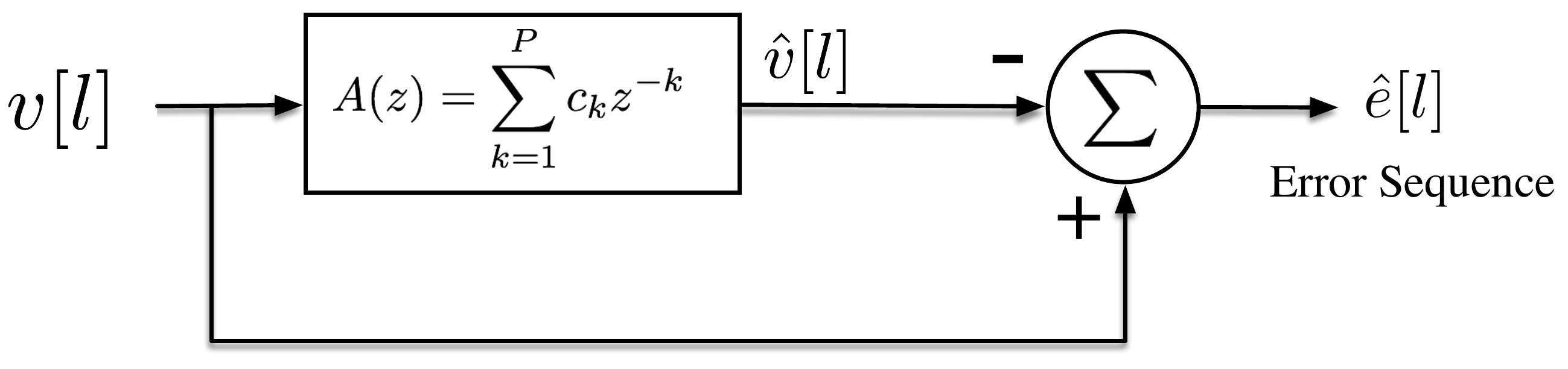}
	\caption{Whitening filter}
	\label{fig:dnfilter}
\end{figure}

The block diagram with the linear predictor used as a whitening filter is shown in Figure \ref{fig:dnfilter}. In the $z$-transform domain the whitening filter $H(z)$ is given by 
\begin{equation}
H(z)=1-A(z)
\end{equation}
where $A(z)$ is the linear prediction filter
\begin{equation}
A(z)=\sum_{k=1}^P c_k z^{-k}.
\end{equation}
The prediction coefficients $c_k$ can be computed using the autocorrelation method of AR modeling. This involves estimating the autocorrelation sequence from the time series and solving the Yule-Walker equations through Levinson Durbin recursion (\citealt{Kay1988,Lju1987}). The design of the whitening filter for each trace can be carried out during an initial noise-only segment of data. Then we whiten each noisy trace before designing the denoising filter $f[\tau]$.

\section{Denoising examples}
\subsection{Synthetic data: Ricker wavelet}
\label{sec:synthDataRicker}
\begin{figure}[htbp]
	\begin{center}
		\includegraphics[width=0.7\textwidth]{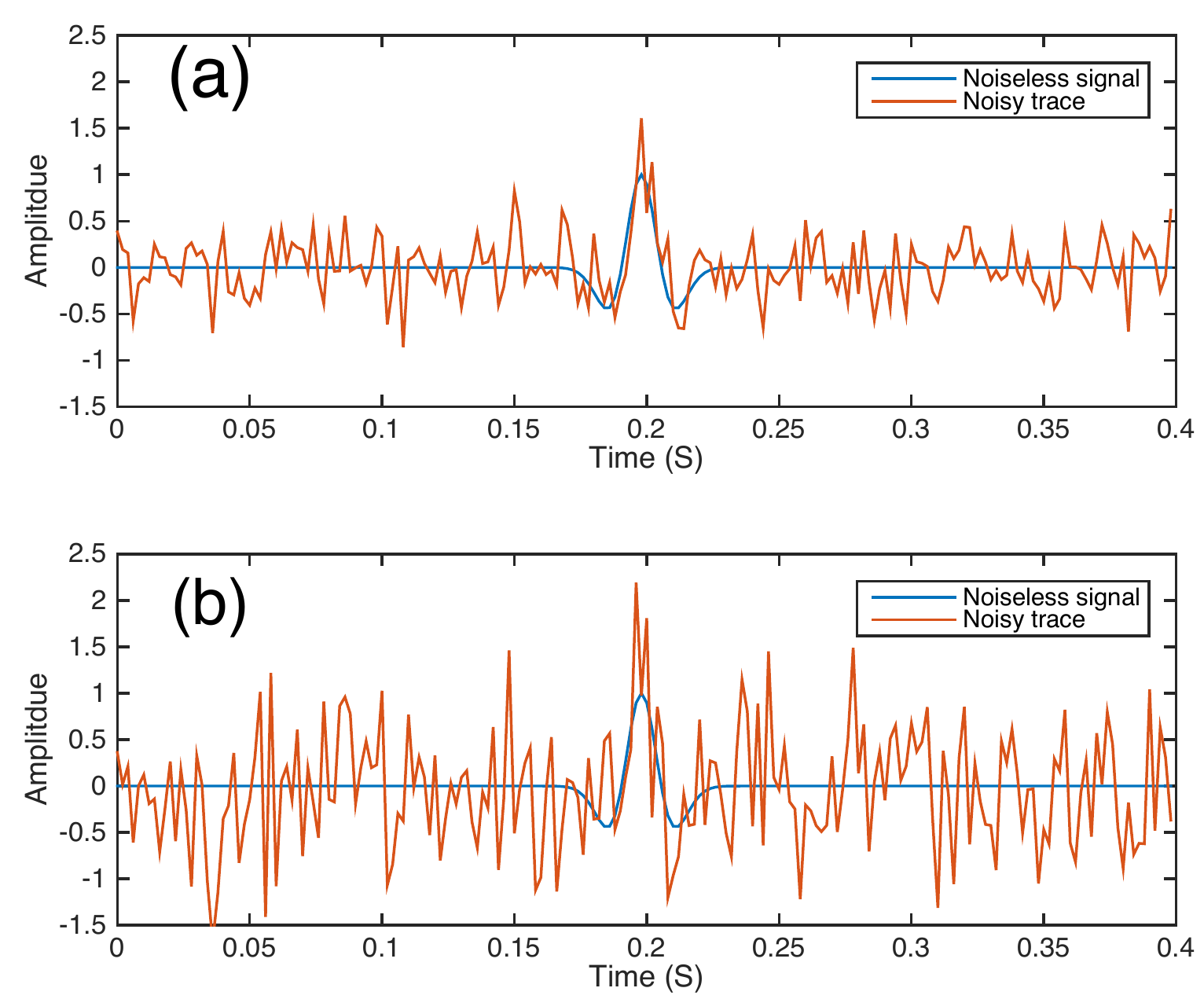}
		\caption{Noiseless and noisy traces, (a) SNR=$-6.03$\,dB; (b) SNR=$-12.01$\,dB. The corresponding PSNR values are $10.46$\,dB and $4.44$\,dB, respectively.
		}
		\label{fig:nVSc}
	\end{center}
\end{figure}

For synthetic data, we assume that there are 200 traces with identical amplitude scaling, i.e.,  $a_i=1$. The waveform $s(t)$ is a Ricker wavelet with center frequency of 30\,Hz and a normalized peak value. The sampling frequency is 500\,Hz and each trace has 200 time samples. 
Based on the SNR given in \eqref{SNR}, two cases are considered: $-6.03$\,dB and $-12.01$\,dB, where the AWGN has $\sigma=0.3$ and $0.6$, respectively. 
In Figure \ref{fig:nVSc} we show noiseless signals superimposed on noisy versions for these two cases.
Recall that plots of the stacked ACFs shown previously in Figure \ref{fig:autoVSfilter} were generated for the case of $\sigma=0.3$.

\begin{figure}[htbp]
	\begin{center}
		\includegraphics[width=0.6\textwidth]{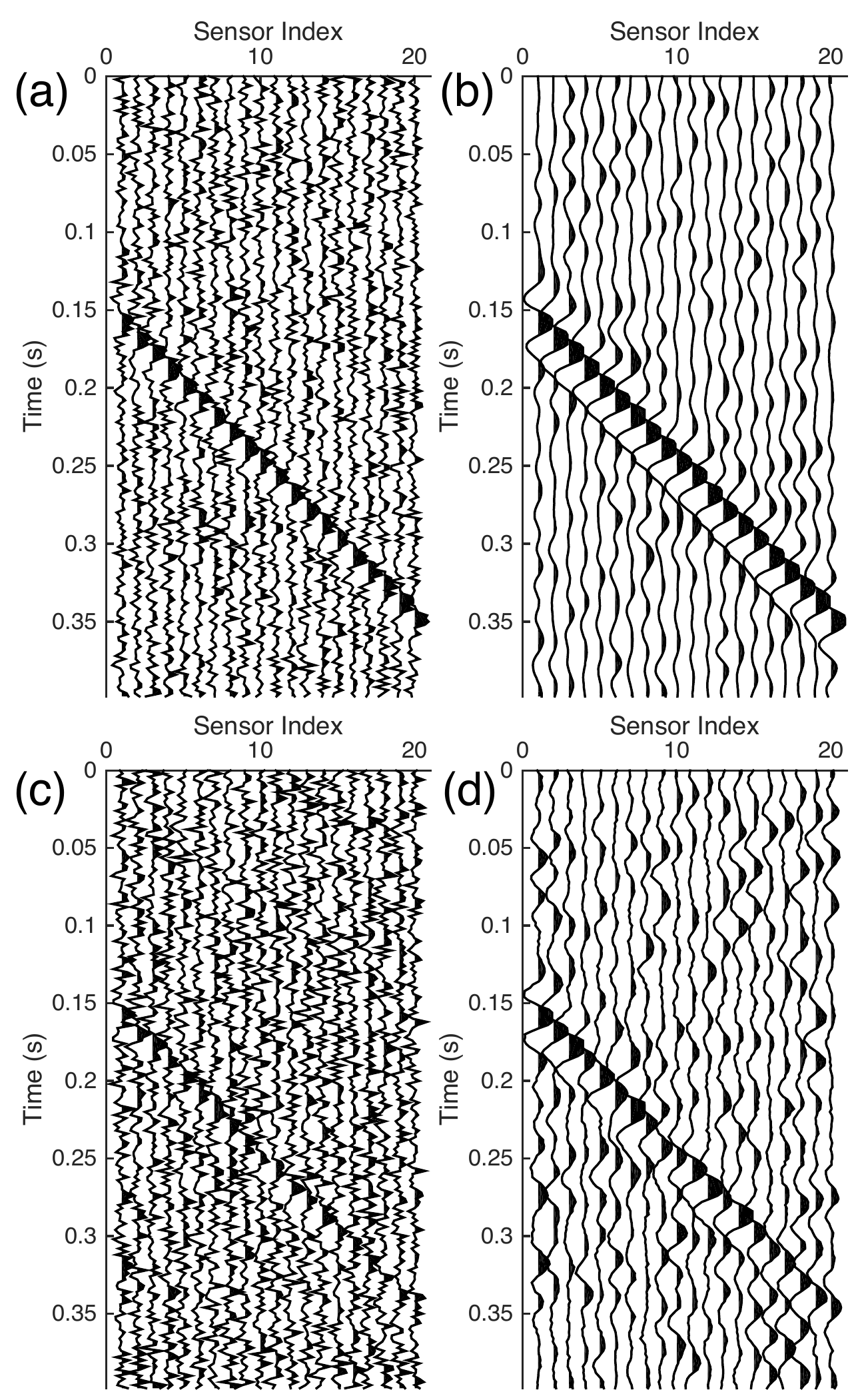}
		\caption{Synthetic data showing 20 out 200 traces. (a) Noisy traces with $\sigma=0.3$, (b) denoising result for the $\sigma=0.3$ case, (c) Noisy traces with $\sigma=0.6$, (d) denoising result for the $\sigma=0.6$ case.}
		\label{fig:DN}
	\end{center}
\end{figure}

For conciseness, we present only the first 20 noisy traces of these two noise levels in Figures \ref{fig:DN}(a) and \ref{fig:DN}(c), respectively. The final denoising results by the proposed scheme are shown in Figures \ref{fig:DN}(b) and \ref{fig:DN}(d). In both cases, the microseismic events in these very low SNR datasets are significantly enhanced. For the low noise case, the denoised data have SNR=2.51\,dB (i.e., PSNR=17.60\,dB). Additionally, the high noise case, the denoised data have SNR=0.51\,dB (i.e., PSNR=10.84\,dB).

\begin{figure}[htbp]
	\begin{center}
		\includegraphics[width=0.7\textwidth]{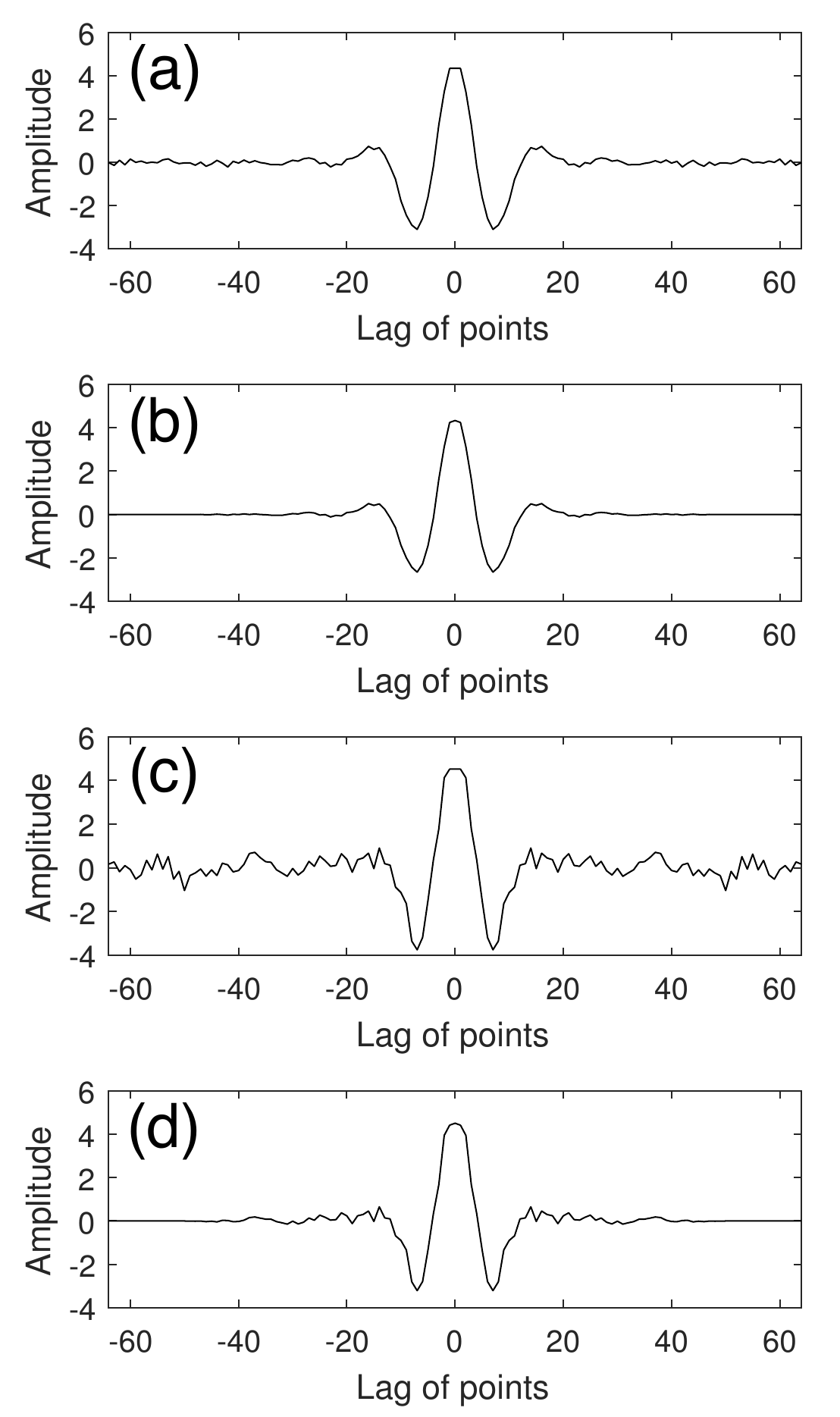}
		\caption{Designed filter for (a) $\sigma=0.3$, without truncation; (b) $\sigma=0.3$, truncated and windowed to $-50\leq\tau\leq 50$; (c) $\sigma=0.6$, without truncation; (d) $\sigma=0.6$, truncated and windowed to $-50\leq\tau\leq 50$.}
		\label{fig:filtercomp}
	\end{center}
\end{figure}

\begin{figure}[htbp]
	\begin{center}
		\includegraphics[width=0.7\textwidth]{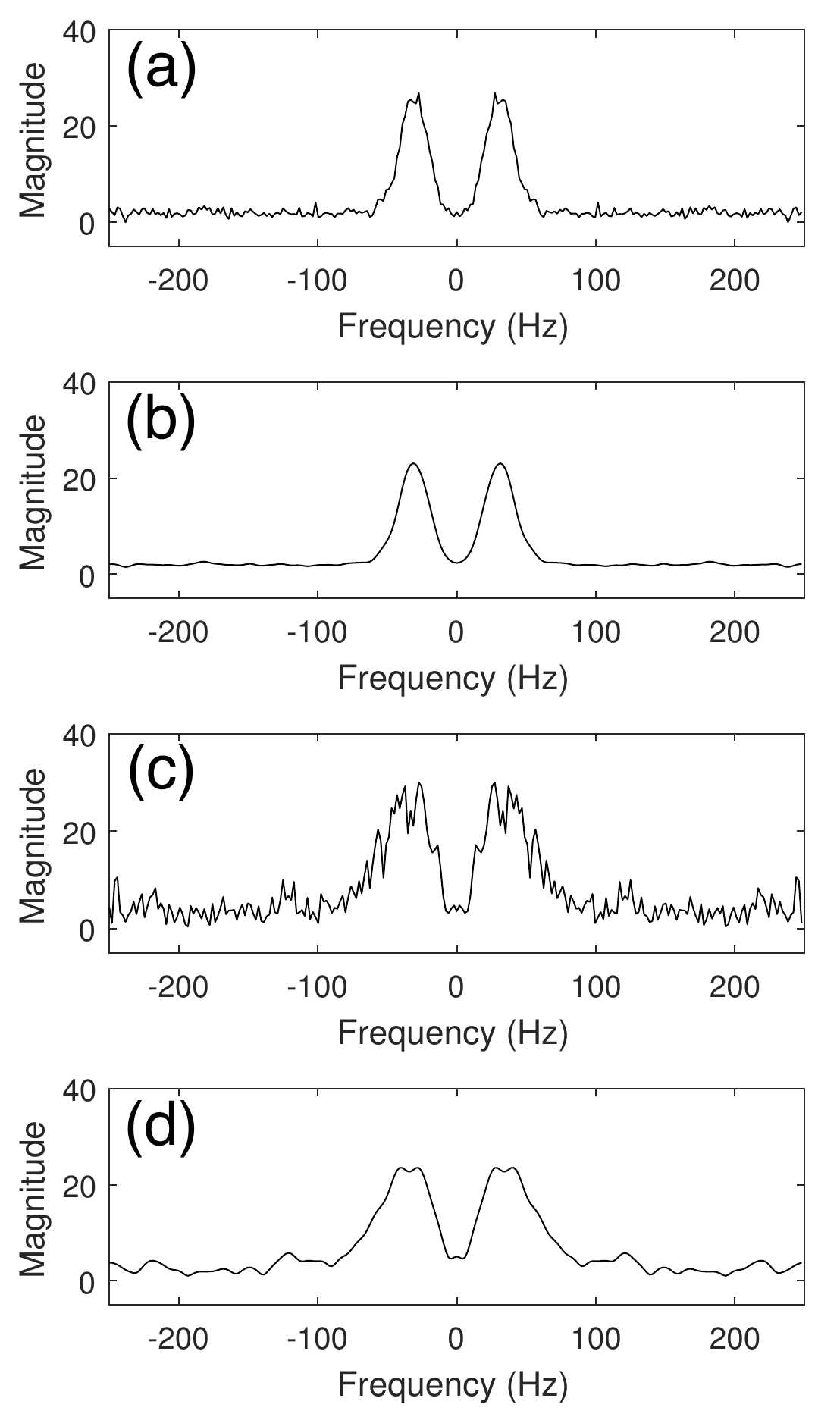}
		\caption{Frequency response of the filters (a) $\sigma=0.3$, without truncation; (b) $\sigma=0.3$, truncated and windowed to $-50\leq\tau\leq 50$; (c) $\sigma=0.6$, without truncation; (d) $\sigma=0.6$, truncated and windowed to $-50\leq\tau\leq 50$.}
		\label{fig:FR}
	\end{center}
\end{figure}

In order to track the intermediate steps of the scheme, we compare the stacked autocorrelation and the designed filter side by side in Figure \ref{fig:autoVSfilter} for the low noise case. Then Figure \ref{fig:filtercomp} illustrates the filter with and without truncation for the $\sigma=0.3$ and $\sigma=0.6$ cases.
The effect of smoothing in the frequency domain is then shown in Figure \ref{fig:FR}, which displays the frequency responses of the filters in Figure \ref{fig:filtercomp}.

\subsection{Field data}
\begin{figure}[htbp]
	\begin{center}
		\includegraphics[width=0.7\textwidth,height=0.75\textwidth]{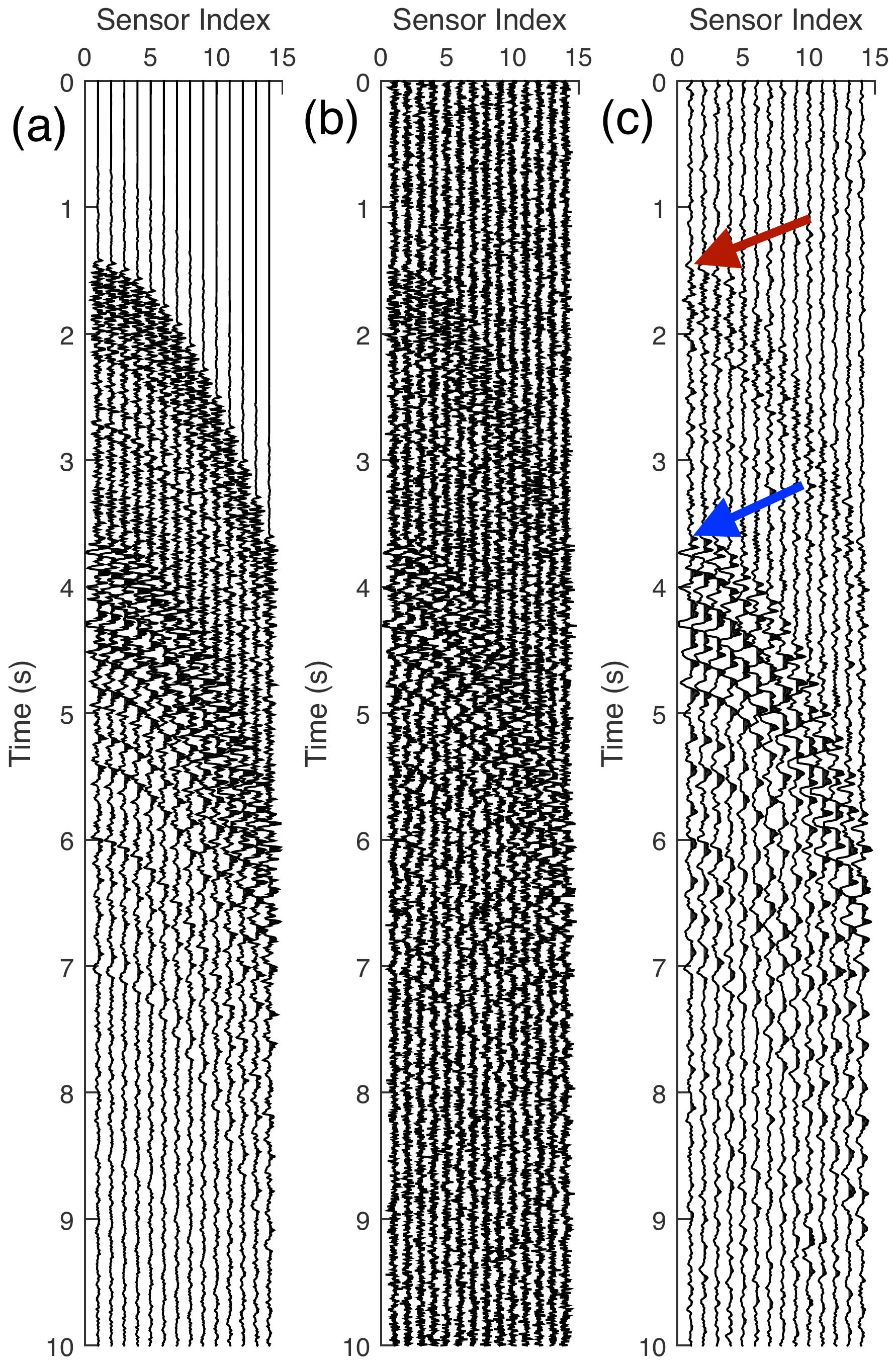}
		\caption{(a) original traces, (b) noisy traces, SNR=$-2.53$\,dB (c) denoising result. P-wave and S-wave arrivals are indicated using red and blue arrows, respectively.}
		\label{fig:nVScFD}
	\end{center}
\end{figure}

\begin{figure}[htbp]
	\begin{center}
		\includegraphics[width=0.7\textwidth]{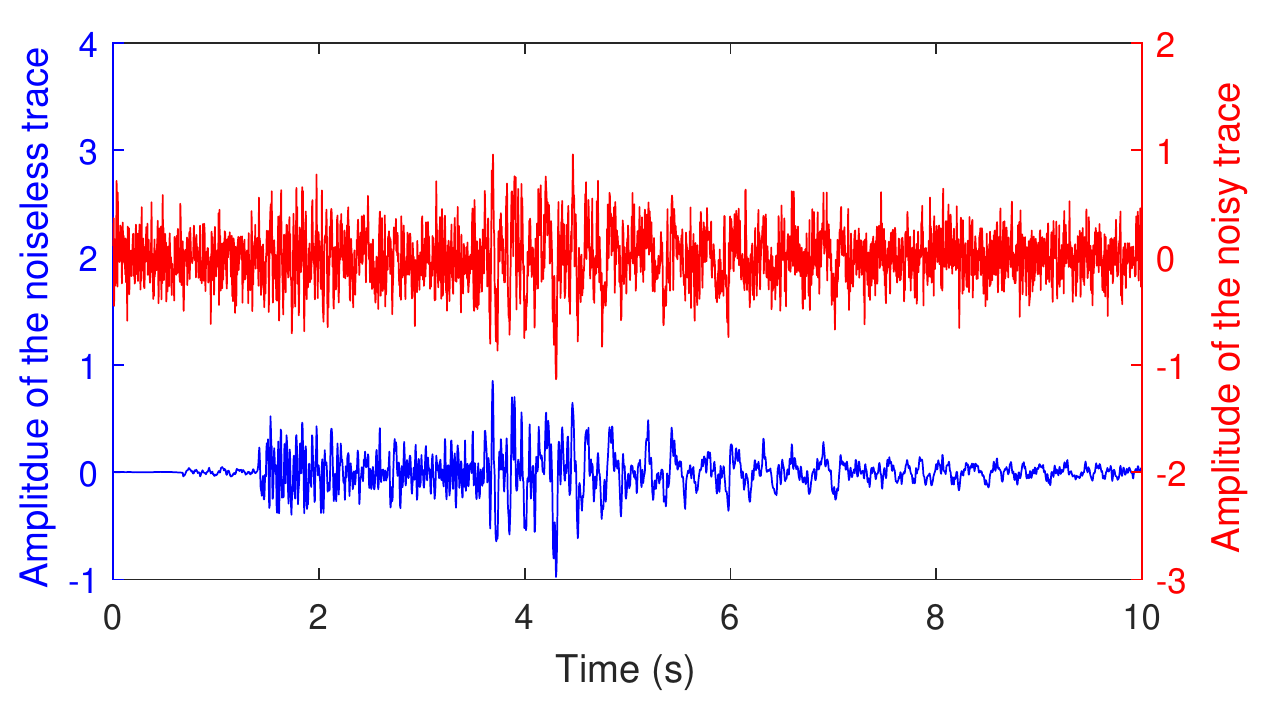}
		\caption{The true waveform (blue) and its noise contaminated version SNR=$-2.53$\,dB (red).}
		\label{fig:noisySampleTrace}
	\end{center}
\end{figure}

In order to test the validity of the proposed scheme with microseismic data that would be received by a surface sensor array, we generate a set of 182 traces with different time delays from a single seismic trace. The waveform used in this study comes from the High Resolution Seismic Network (HRSN) operated by Berkeley Seismological Laboratory, University of California, Berkeley, the Northern California Seismic Network (NCSN) operated by the U.S. Geological Survey, Menlo Park, and distributed by the Northern California Earthquake Data Center (NCEDC). The sampling frequency is 250\,Hz and each trace contains data for 10 seconds. We rescale every trace to have a normalized peak value of 1.0 and regard them as clean data. We then add AWGN of $\sigma=0.2$ to the clean data, which gives SNR=$-2.53$\,dB (i.e., PSNR=$13.98$\,dB). For conciseness we only show 14 clean and noisy traces in Figures \ref{fig:nVScFD}(a) and \ref{fig:nVScFD}(b). In Figure \ref{fig:noisySampleTrace} we present a clean sample trace and its noisy version as a close-up.

\begin{figure}[htbp]
	\begin{center}
		\centering
		\includegraphics[width=0.7\linewidth]{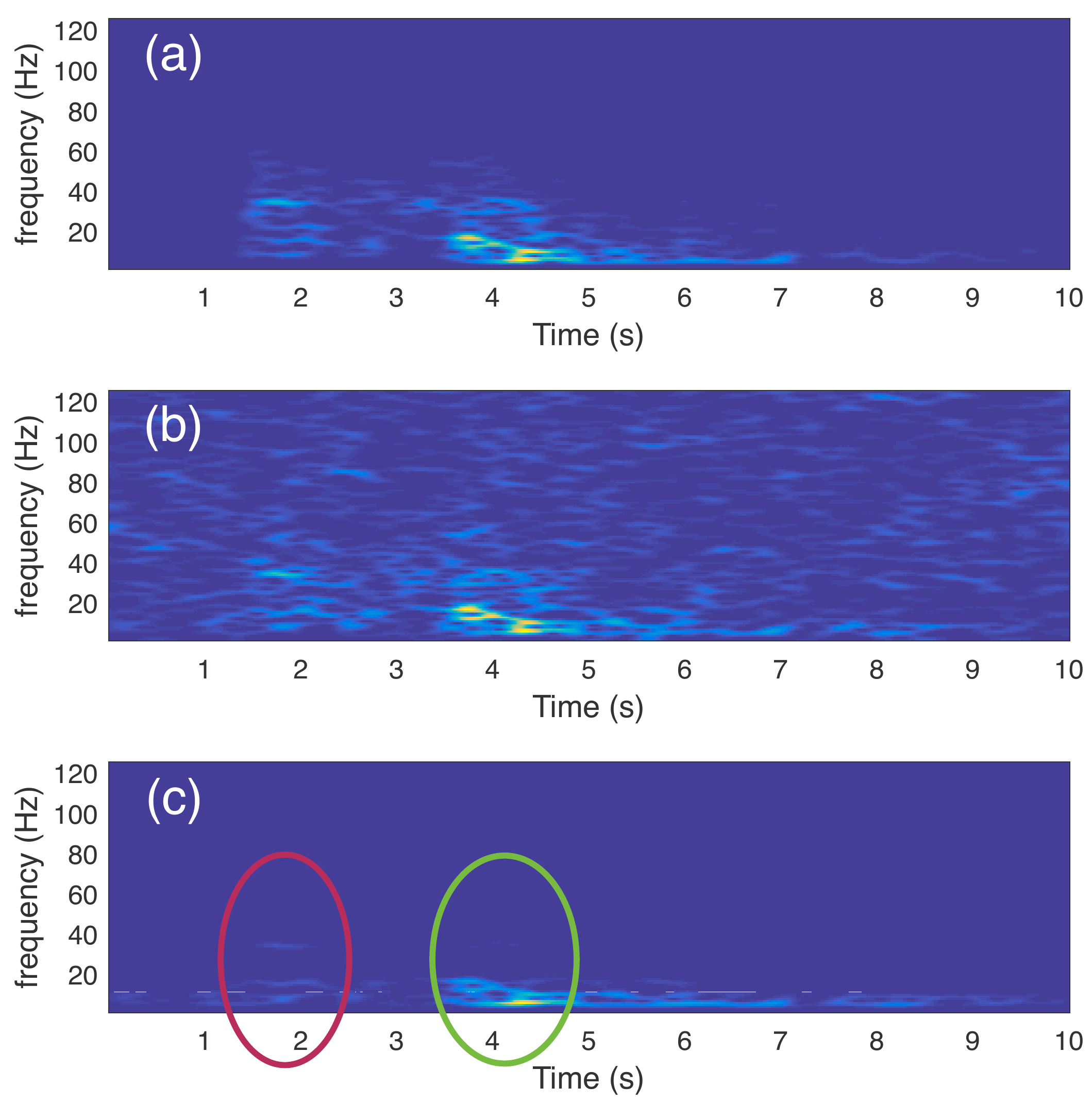}
		\captionof{figure}{Spectrogram of (a) original trace, (b) noisy trace, SNR=$-2.53$\,dB, and (c) denoising result. The red and green circles indicate the P- and S-wave arrivals, respectively.}
		\label{fig:spect}
	\end{center}
\end{figure}

\begin{figure}[htbp]
	\begin{center}
		\centering
		\includegraphics[width=0.9\linewidth]{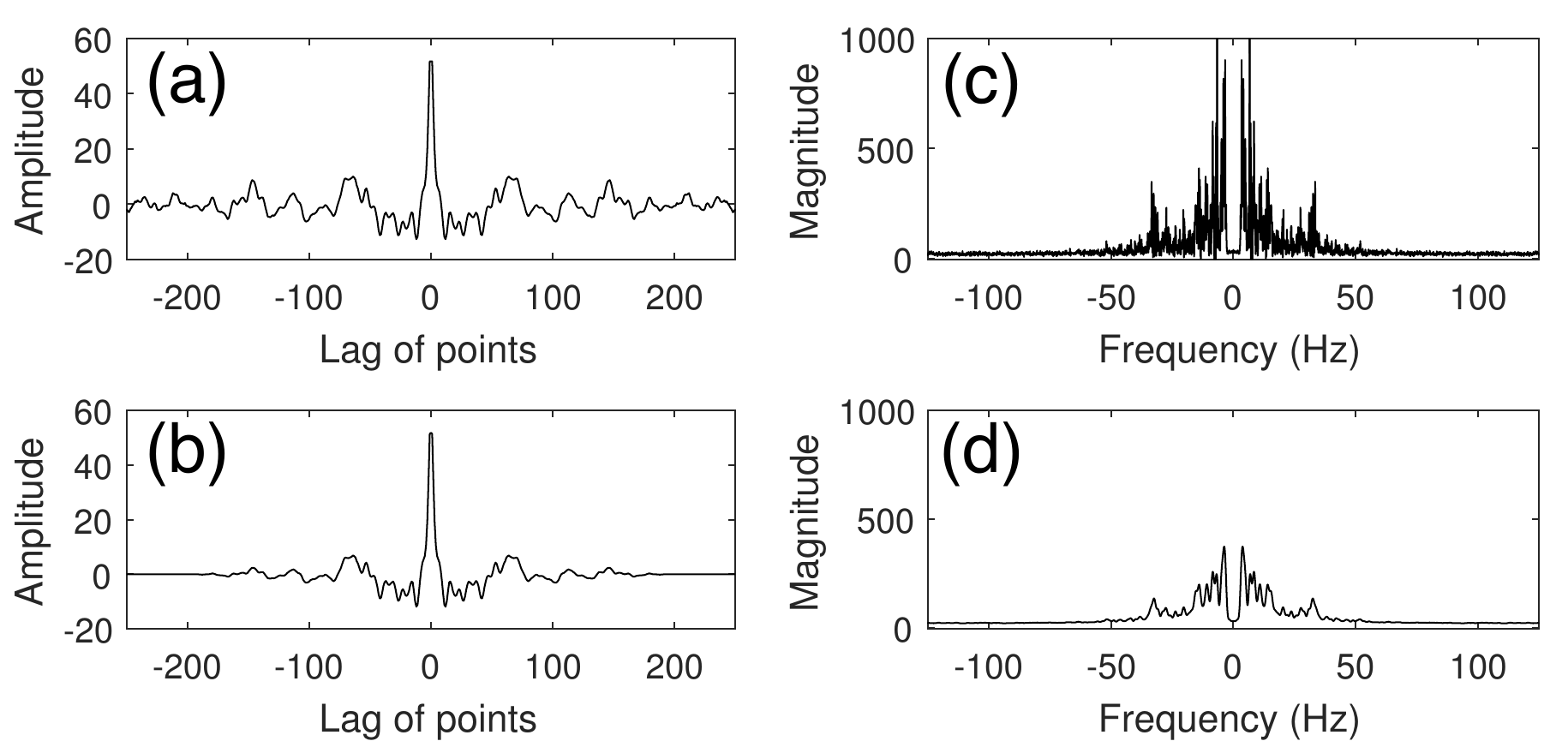}
		\captionof{figure}{Impulse response of ACF-designed filter (a) without truncation, (b) truncated and windowed to $-190\leq\tau\leq190$. Frequency response (c) without truncation, (d) truncated and windowed to  $-190\leq\tau\leq190$.}
		\label{fig:FilterCompFD}
	\end{center}
\end{figure}

The denoising result using the ACF-designed filter with a truncation window of $\tau=190$ is shown in Figure \ref{fig:nVScFD}(c). We not only observe that the denoising scheme clearly recovers the P-wave and S-wave arrivals (indicated using red and blue arrows, respectively) but also well preserves the waveform from noisy traces where precise manual detection of the signal is almost impossible. The denoising result for a sample trace is shown in Figure \ref{fig:spect}, where we note the P-wave and S-wave arrivals are preserved in the regions indicated by red and green circles, respectively, on the spectrogram in Figure \ref{fig:spect}(c). As reference, we present the filter with and without truncation and the corresponding frequency response amplitude in Figure \ref{fig:FilterCompFD}, where the smoothing effect is easy to see.

The methodology described above was tested on a real data set. However, due to the proprietary nature of the data set we are unable to include those results. This is the reason why we used the simulation in Figure \ref{fig:nVScFD} that is generated by manually shifting real traces. As we indicated earlier, the prerequisite for this method to work is that the recorded data across all sensors in the array must have similar frequency content (although the waveforms may be altered slightly). With the proprietary real data set, we obtain results similar to what is shown in Figure \ref{fig:nVScFD} above.
	
	Finally, note that this method is effective in suppressing uncorrelated noise, not correlated (i.e., colored) noise.
	Consequently, we do not expect our method to be applicable in all noise scenarios. In the next section we show that some preconditioning can be performed to decorrelate additive noise when prior knowledge is available.

\subsection{Examples for noise prewhitening}
\begin{figure}[htbp]
	\begin{center}
		\includegraphics[width=0.7\textwidth]{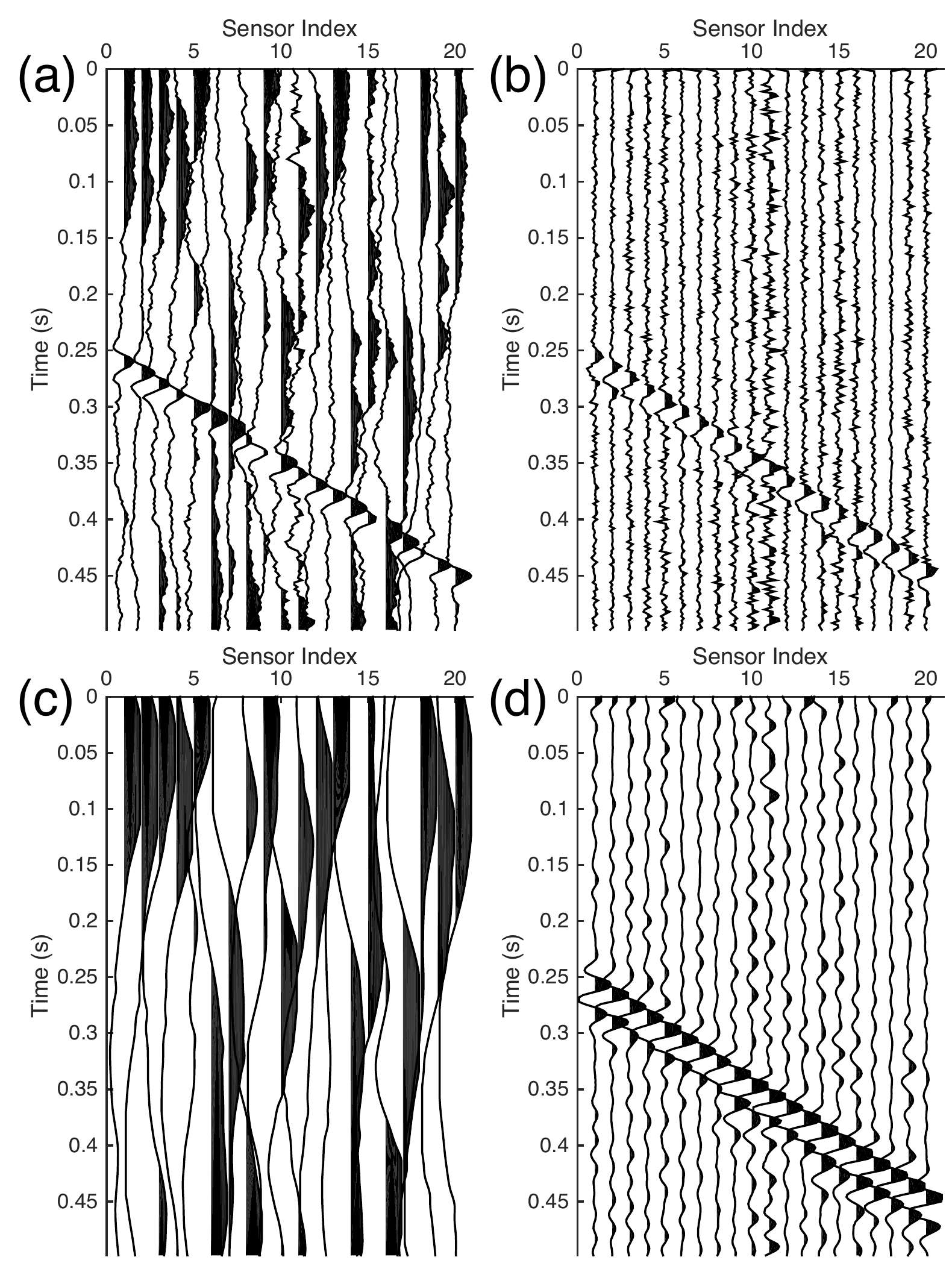}
		\caption{Denoising results for colored noise. (a) noisy data, colored noise SNR=$-10$\,dB,  (b) data after applying a 20th order prewhitening filter, (c) denoising without prewhitening, (d) denoising result after prewhitening.}
		\label{fig:PreWhitDN}
	\end{center}
\end{figure}

In order to verify the necessity and validity of prewhitening for microseismic data with additive colored noise, we consider 200 microseismic noisy traces of SNR=$-10$\,dB (the PSNR is not calculated since the noise is not AWGN). The first 20 noisy traces are shown in Figure \ref{fig:PreWhitDN}(a). The power spectrum of the noise is higher for low frequency and is identical for all sensors. A prewhitening filter of degree 20 is learned from the signal segments that contain only noise. The output of the prewhitening filter is shown in Figure \ref{fig:PreWhitDN}(b). Figure \ref{fig:PreWhitDN}(d) shows the superiority of the prewhitened and denoised result to denoising without prewhitening in Figure \ref{fig:PreWhitDN}(c).

\begin{figure}[htbp]
	\begin{center}
		\includegraphics[width=0.7\textwidth]{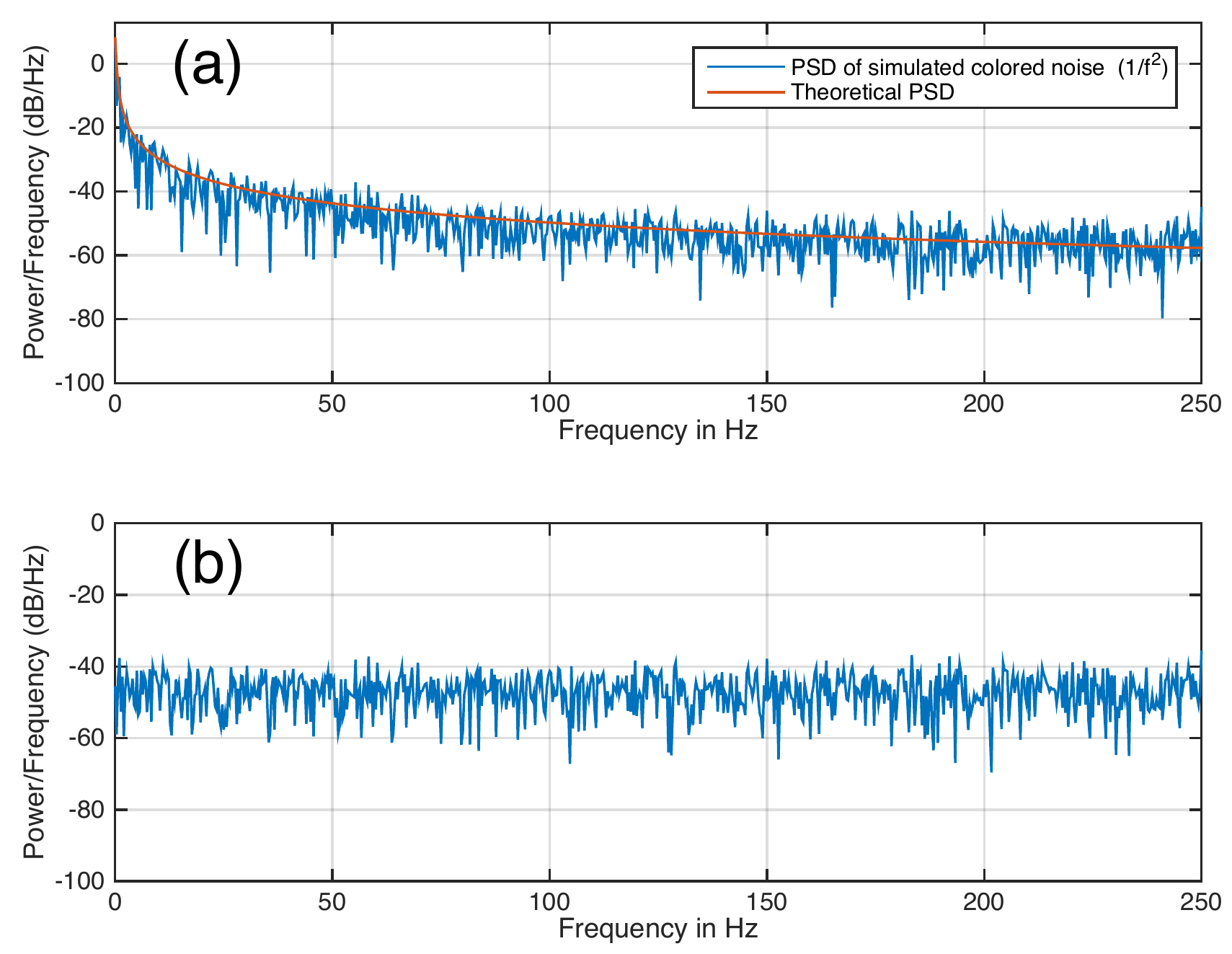}
		\caption{Power spectrum density (PSD) of (a) colored noise, and (b) after prewhitening.}
		\label{fig:PowerPreWhite}
	\end{center}
\end{figure}

In order to demonstrate the validity of the prewhitening process, we compare the noise power spectrum of noise-only segment on the first sensor before and after whitening (see Figure \ref{fig:PowerPreWhite}). We can see the power spectrum of the noise is effectively flattened.

\section{Detection}
In order to alleviate the computational burden when processing large data sets acquired with passive monitoring, it is always desirable to have an accurate detection of seismic events as the first component of the microseismic data processing pipeline. The total work load can be significantly reduced, if we apply further processing only to the part of data that contains detected seismic events. Conventional detection schemes are commonly based on changes in certain characteristics along single traces, such as trace energy, absolute value of amplitude, short-term-average/long-term-average (STA/LTA) type algorithms (\citealt{EarShe1994}), and waveform correlation with strong events (\citealt{GibRin2006,MicTok2007,SonKulAO2010}), to name a few. However, these methods require either good SNR or isolated strong events that serve as a template. Otherwise, they typically produce many false alarms when an aggressive threshold is employed to detect small events. 

In the foregoing sections, we have shown that the stacked autocorrelation can efficiently estimate the frequency characteristics of a common waveform in the presence of AWGN for multi-channel microseismic data. If waveforms originating from the same seismic event are received across the array of sensors, high coherence is commonly observed. Based on this fact, we devise a multi-channel detector that is capable of exploiting the information on all channels at once. 

In order to achieve a reasonable detection resolution in the time domain, we adopt a sliding window technique from the conventional schemes. The detection indicator $\eta(t)$, which corresponds to the sliding window starting at $t$, is defined as follows:
\begin{equation}\label{eta}
\eta(t)=\frac{1}{N}\sum_{i=1}^N \frac{\max_\omega |\cF\{  x_i \star x_i \}|}{\|\cF\{  x_i \star x_i \} \|_1},
\end{equation}
where $x_i(t)$ is the $i$-th trace truncated (and weighted) by the sliding window starting at $t$. Since the denominator $\|\mathcal F\{x_i\star x_i\}\|_1$ is proportional to the energy of $x_i$, each trace is normalized prior to summation in \eqref{eta}.
Therefore, the detector defined in \eqref{eta} measures only the resemblance of the traces which is independent of their amplitudes.
In data from natural micro-earthquakes, or microseismic data from hydraulic fracturing, the amplitudes of signals could vary significantly across a sensor array.
Thus, the normalization in \eqref{eta} for all traces is necessary.


\section{Detection example}
\begin{figure}[htbp]
	\centering
	\begin{minipage}{0.7\textwidth}
		\centering
		\includegraphics[height=8cm]{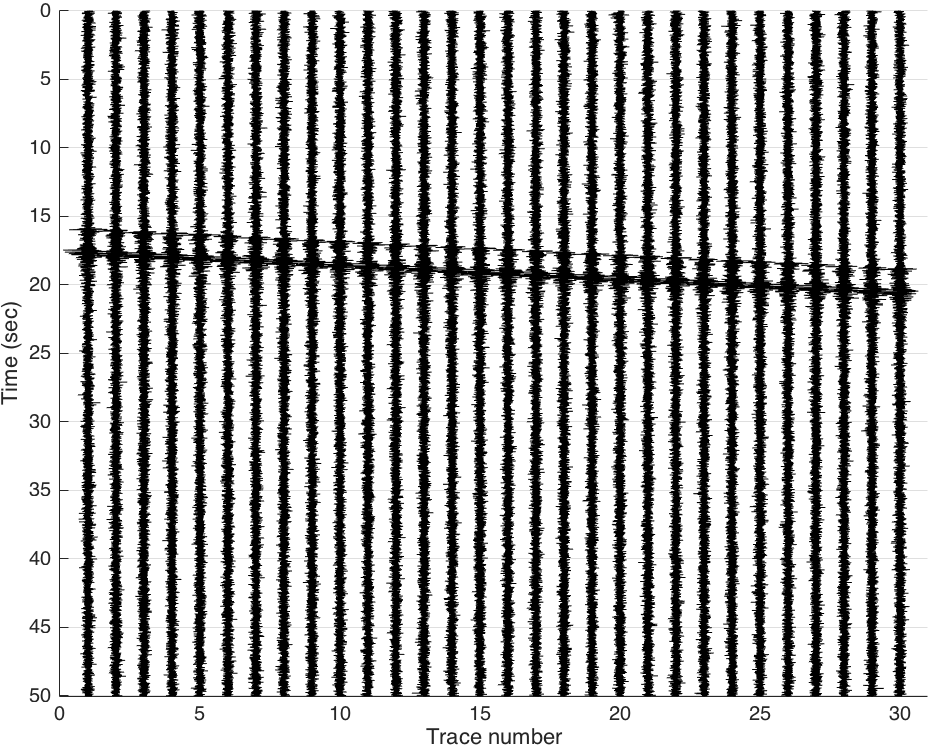}\\
		(a) Synthetic traces, $\sigma =0.1$ (PSNR = 20\,dB)
	\end{minipage}
	\begin{minipage}{0.28\textwidth}
		\centering
		\includegraphics[height=8cm]{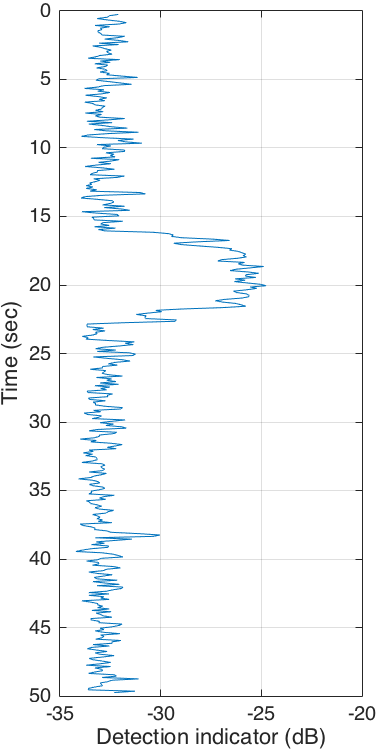}\\
		(b) Detection indicator
	\end{minipage}\\
	~\\[3ex]
	\begin{minipage}{0.7\textwidth}
		\centering
		\includegraphics[height=8cm]{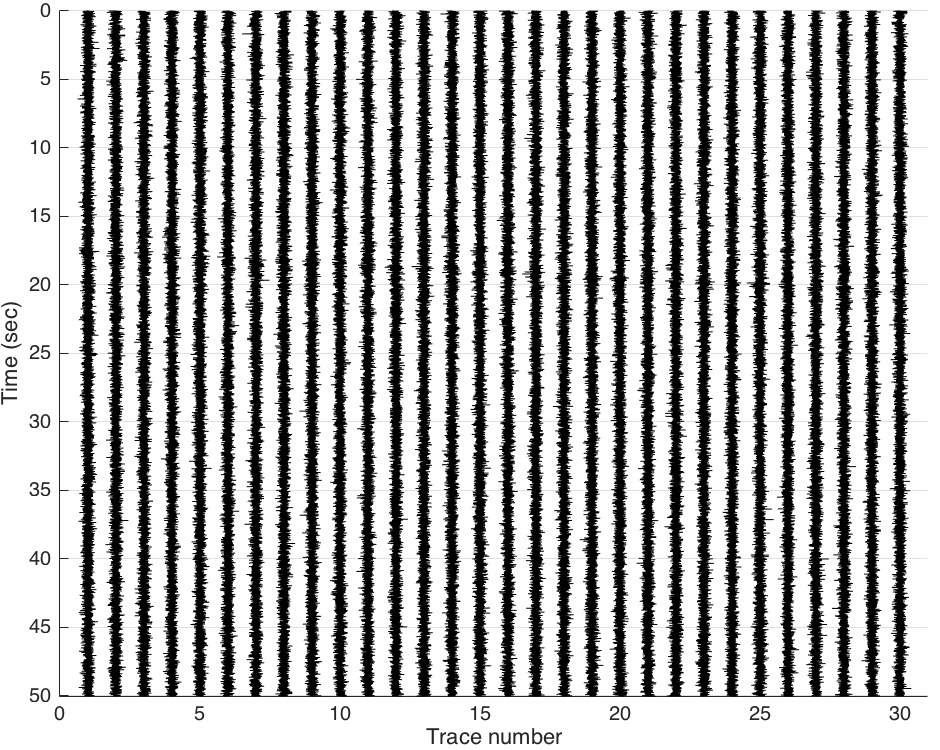}\\
		(c) Synthetic traces, $\sigma = 0.4$ (PSNR = 8\,dB)
	\end{minipage}
	\begin{minipage}{0.28\textwidth}
		\centering
		\includegraphics[height=8cm]{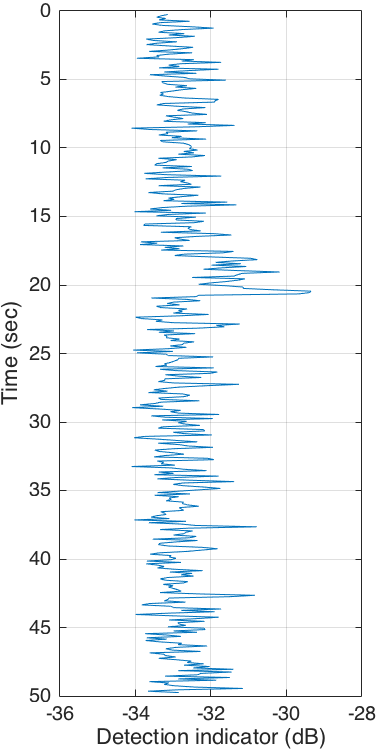}\\
		(d) Detection indicator
	\end{minipage}
	\caption{Synthetic multichannel data using 30 real seismic traces with linear moveout totaling 5 seconds contaminated by AWGN of (a) $\sigma=0.1$  and (c) $\sigma=0.4$. Pre-detection results using the indicator \eqref{eta} are shown in (b) and (d), respectively.}
	\label{fig:sync_example}
\end{figure}

In this section, a synthetic example using a seismic data section obtained by manually time-shifting a real seismic trace plus random noise is shown in Figure~\ref{fig:sync_example} to demonstrate the effectiveness of the pre-detection indicator $\eta(t)$ based on the stacked autocorrelation as proposed in \eqref{eta}. The synthetic data shown in Figures~\ref{fig:sync_example}(a) and \ref{fig:sync_example}(c) includes 30 traces which were delayed with a linear moveout between 15 and 20 seconds from a single real seismic trace which consists of both P-wave and S-wave phase. The seismic trace is from the same data set as Figure \textcolor{blue}{\eqref{fig:nVScFD}(a)}, whose sampling frequency is 250\,Hz. Additive white Gaussian noise of $\sigma=0.1$ (for high SNR at PSNR = 20\,dB) and $\sigma=0.4$ (for low SNR at 8\,dB) is used in Figures~\ref{fig:sync_example}(a) and \ref{fig:sync_example}(c), respectively. We use a sliding window of length 0.5\,sec with an overlap of 0.4\,sec and then compute $\eta(t)$ with a 128-point FFT.

In the high SNR case, the detection indicator $\eta(t)$ returns obvious high values during the time frames of the coherent signal arrival, while the rest of the time seems to exhibit a noise floor at about $-33$\,dB. Setting a threshold at $-29$\,dB would give nearly perfect detection of the simulated microseismic events in the time domain. In the low SNR case, the noise floor stays at about the same value, but the detection region has much smaller values and it would be more difficult to set a threshold to separate the true arrivals from the noise floor. Since $\eta(t)$ measures the coherence among traces, the noise floor does not change with the additive noise amplitudes. Thus, as the noise amplitudes increase, the $\eta(t)$ values within the coherent signal region decrease and will eventually fall below the noise floor. As demonstrated in Figure~\ref{fig:sync_example}(d), our proposed detection indicator $\eta(t)$ can deal with a PSNR as low as 8\,dB using hard thresholding. Furthermore, improving the peak detector with methods such as smoothing or polynomial fitting can further reduce the lower-bound of detection PSNR.

%

\section{Conclusion}
Surface microseismic data, which is typically noisy, requires  robust detection and an explicit denoising step before further processing. We present a multi-channel denoising and detection method based on autocorrelations which can effectively suppress uncorrelated noise without knowing relative time offsets. 
A prewhitening scheme extends the applicability of this denoising filter to more general and practical scenarios of microseismic monitoring.
The effectiveness of the detection, denoising scheme, and the prewhitening for colored noise is tested using synthetic and real seismic waveforms.  

\section*{Acknowledgement}
This work is supported by the Center for Energy and Geo Processing
at Georgia Tech and King Fahd University of Petroleum and
Minerals. Authors appreciate KFUPM support of this work under grant number GTEC1311.


\end{document}